\documentclass[a4paper,12pt]{article}
\pdfoutput=1
\usepackage{graphicx}

\newcommand{\bmat}{\left(\begin{array}}
\newcommand{\emat}{\end{array}\right)}

\def\beq{\begin{equation}}
\def\eeq{\end{equation}}
\def\beqa{\begin{eqnarray}}
\def\eeqa{\end{eqnarray}}

\def\-{\hphantom{-}}

\def\s2{\frac{1}{\sqrt2}}

\def\beq{\begin{equation}}
\def\eeq{\end{equation}}
\def\beqa{\begin{eqnarray}}
\def\eeqa{\end{eqnarray}}
\def\ba{\begin{array}}
\def\ea{\end{array}}

\def\IF{\relax{\rm I\kern-.18em F}}
\def\II{\relax{\rm I\kern-.18em I}}
\def\IP{\relax{\rm I\kern-.18em P}}
\def\IC{\relax\hbox{\kern.25em$\inbar\kern-.3em{\rm C}}}
\def\IR{\relax{\rm I\kern-.18em R}}

\def\cp{{\cal P}}

\def\Dsl{\,\raise.15ex\hbox{/}\mkern-13.5mu D} 
\def\IZ{Z\kern-.4em  Z}

 \def\cp#1{\relax\ifmmode {\IP\kern-2pt{}_{#1}}\else $\IP\kern-2pt{}_{#1}$\=fi}

\catcode`\@=11
\newdimen\@rotdimen
\newbox\@rotbox

\def\@vspec#1{\special{ps:#1}}
\def\@rotstart#1{\@vspec{gsave currentpoint currentpoint translate
   #1 neg exch neg exch translate}}
\def\@rotfinish{\@vspec{currentpoint grestore moveto}}
\def\@rotr#1{\@rotdimen=\ht#1\advance\@rotdimen by\dp#1%
   \hbox to\@rotdimen{\hskip\ht#1\vbox to\wd#1{\@rotstart{90 rotate}%
   \box#1\vss}\hss}\@rotfinish}
\def\@rotl#1{\@rotdimen=\ht#1\advance\@rotdimen by\dp#1%
   \hbox to\@rotdimen{\vbox to\wd#1{\vskip\wd#1\@rotstart{270 rotate}%
   \box#1\vss}\hss}\@rotfinish}%
\def\@rotu#1{\@rotdimen=\ht#1\advance\@rotdimen by\dp#1%
   \hbox to\wd#1{\hskip\wd#1\vbox to\@rotdimen{\vskip\@rotdimen
   \@rotstart{-1 dup scale}\box#1\vss}\hss}\@rotfinish}%
\def\@rotf#1{\hbox to\wd#1{\hskip\wd#1\@rotstart{-1 1 scale}%
   \box#1\hss}\@rotfinish}%
\def\rotate{\@ifnextchar[{\@rotate}{\@rotate[l]}}
\def\@rotate[#1]#2{\setbox\@rotbox=\hbox{#2}\@nameuse{@rot#1}\@rotbox}

\catcode`\@=12

\begin{document}

\makeatletter \@addtoreset{equation}{section} \makeatother
\renewcommand{\theequation}{\thesection.\arabic{equation}}
\pagestyle{empty}
\pagestyle{empty}
\rightline{\today}
\vspace{1.5cm}
\setcounter{footnote}{0}


\begin{center}
\renewcommand{\thefootnote}{\fnsymbol{footnote}}
\LARGE{\bf  Scaling Properties of the Lorenz System \\ 
and  Dissipative Nambu Mechanics\footnote[2]{The present work is dedicated to the memory
of Prof. J.S.Nicolis}}
\renewcommand{\thefootnote}{\arabic{footnote}}\\[2mm]
{\large{\bf Minos Axenides$^{1}$
 and Emmanuel Floratos$^{1,2}$}}
\\[4mm]
 \normalsize{\em $^1$ Institute of Nuclear and Particle Physics, N.C.S.R. Demokritos,\\
GR-15310, Agia Paraskevi, Attiki, Greece}\\
\normalsize{\em $^2$  Department of Physics, Univ. of Athens,
GR-15771 Athens, Greece}\\[5mm]  axenides@inp.demokritos.gr; \ \ \   mflorato@phys.uoa.gr
\end{center}
\vspace{2.0mm}
\begin{center}
{\large {\bf Abstract}}
\end{center}

\vspace{2.0mm}
{\small  
In the framework of Nambu Mechanics, we have recently argued that Non-Hamiltonian Chaotic Flows in $ R^{3} $, are dissipation induced deformations, of integrable volume preserving flows,  specified by pairs of Intersecting Surfaces in $R^{3}$.  
In the present work we focus our attention to the Lorenz system with a linear dissipative sector in its phase space dynamics. In this case the 
Intersecting Surfaces are Quadratic. We parametrize its dissipation strength through a continuous control  parameter  $\epsilon$, acting homogeneously over the whole 3-dim. phase space. 
In the extended $\epsilon$-Lorenz system we find a scaling relation between the dissipation strength $ \epsilon $
and Reynolds number  parameter r . It results from the scale covariance, we impose on the Lorenz equations
under arbitrary rescalings of all its dynamical coordinates. 
Its integrable limit, ($ \epsilon = 0 $, \ fixed  r), which is described in terms of intersecting Quadratic  
 Nambu ``Hamiltonians" Surfaces, gets mapped on the infinite value limit of the Reynolds number parameter \ (  r $\rightarrow \infty,\ \epsilon= 1$).
 In effect weak dissipation, through small $\epsilon$ values, generates and controls the  well explored Route to Chaos in the large r-value regime. The non-dissipative $\epsilon=0 $ integrable limit is therefore the gateway  to Chaos for the Lorenz system.



\newpage

\setcounter{page}{1} \pagestyle{plain}
\renewcommand{\thefootnote}{\arabic{footnote}}
\setcounter{footnote}{0}
\tableofcontents

\section{Introduction}

In the framework of Nambu mechanics we have recently\cite{af1,af2} proposed a geometric approach for the study of dynamical dissipative systems \cite{chaos}. It is implemented through a special
decomposition of the flow of the system in two parts. The non-dissipative and threfore phase-space volume preserving 
component and the dissipative constituent part which is volume contracting. In the particular case of $ D=3 $ phase-space dimensions such a splitting takes the suggestive form

\beq
\dot{ \vec{x}} \ = \ \vec{\nabla} H_{1} \ \times \vec{\nabla}H_{2} \ + \ \vec{\nabla} D 
\eeq
where $H_{1}, H_{2} $ are the Nambu ``Hamiltonians'' or equivalently, the Clebsch-Monge potentials \cite{nambu} where D is the dissipation potential. In particular, for the Lorenz
system \cite{lorenz,sparrow}, 

\beqa
\dot{x} &=& \sigma(y-x) \nonumber \\ \dot{y} &=& x(r-z) - y \\ \dot{z} &=& x y - bz \nonumber
\eeqa   

the functions $H_{1}, H_{2}$,D are:

\beqa
H_{1} &=& \frac{1}{2} \left(y^{2} + ( z - r )^{2} \right) \nonumber \\
H_{2} &=& \sigma z - \frac{x^{2}}{2} \\ 
D &=& - \ \frac{1}{2} \ (\sigma x^{2} + y^{2} +b z^{2} ) \nonumber
\eeqa

The non-dissipative Lorenz system which was studied carefully in $\cite{af1}$  was also investigated some time ago in  \cite{nevir} and in a similar vein further back in time by \cite{haken}

\beq
\dot{ \vec{x}} = \vec{\nabla} H_{1} \times \vec{\nabla}H_{2} 
\eeq

is an integrable dynamical system, where the trajectories in phase space are given by the intersection
of the surfaces defined by the conserved Nambu `` Hamiltonians'' $ (H_{1}, H_{2})$ . In this picture the dynamical system
in
rel(1.1) is defined through its non-dissipative part of eq,(1.4), over which  its phase space dissipation  $ \vec{\nabla} D$ operates.
In order to control its dissipation strength and take hold as accurately as possible of its phase 
space volume contracting effect, we introduce an associated control parameter $ \epsilon $
as follows:
\beq
\dot{\vec{x}} = \vec{\nabla} H_{1} \times \vec{\nabla}H_{2} \ + \ \epsilon \vec{\nabla} D
\eeq
 It is defined in a range of values on the interval $[0,1]$ with its value $\epsilon=0$ recovering the Lorenz system nondissipative part, whereas for $ \epsilon=1$ the full system being  obtained. Clearly the $\epsilon$-parameter controls
the dissipation rate of the phase-space volume of the dynamical system  in question. For the Lorenz case, for example,

\beq
\delta V(t) \ = \ e^{- \epsilon(1 + \sigma + b ) t} \ \delta V(0)
\eeq

We could say that this is nothing else but a rescaling of time. As we shall demonstrate in the particular but very interesting
case of the Lorenz system,  the rescaling of time is possible only under appropriate simultaneous rescalings of all the
dynamical variables x,y,z of the phase space, the time t and more importantly, of the Reynolds number r , while keeping the two others b, $\sigma$ intact. 

At this stage we could observe that the introduction of the control parameter $\epsilon$ measures also the strength of the attraction of the Lorenz ellipsoid \cite{lorenz}.

Indeed, the rate of change of the Liapunov function
\beq
V= \frac{1}{2} \left( r x^{2} + \sigma y^{2} + \sigma (z - 2 r )^{2}\right) = 
\sigma H_{1} + r H_{2} + \frac{3}{2} r^{2} \sigma
\eeq

is given by

\beq
\dot{V} = \dot{\vec{r}} \cdot \nabla V = - \epsilon \sigma \left[ r x^{2} + y^{2} +
b (z - r)^{2} - b r^{2} \right] 
\eeq

When $ \epsilon$ is a small positive number, the velocity is entering the Lorenz
ellipsoid V, if all of its points are located outside the ellipsoid defined by the vanishing  of the expression of the last equation. On the other hand, in the case of the nondissipative system of eq.(1.4), i.e.
 $\epsilon=0 $ the velocity of the particle is tangent to the ellipsoid 
 $ V=V(x_{0},y_{0},z_{0}) $ where $(x_{0},y_{0},z_{0})$ are the initial conditions of the motion, because $H_{1}, H_{2}$ are conserved. Thus $\epsilon$ controls the overall 
 attractive strength  of the attractor.
 
In our recent work we also introduced a method of Matrix-Heisenberg  Quantization for the Lorenz system aiming to examine the compatibility
of the classical Lorenz strange attractor dynamics with the fundamental principles of Quantum Mechanics. We defined rigorously
the quantization procedure starting from the non-dissipative system (1.4) where we replaced the phase space coordinates x,y,z 
with $ N \times N $ hermitean matrices $\hat{X} , \hat{Y}, \hat{Z}$  consistently with the appropriate rules of Quantum Correspondence. The quantum dissipation  was defined to respect the above quantum
correspondence principle. It is intuitively obvious that, if dissipation strength in phase space is big enough, the approach 
of the system to the classical limit is fast and efficient expressed through  the exponentially fast vanishing in time of the commutators of   
$\hat{X} , \hat{Y}, \hat{Z}$. Thus it is imperative for the study of the coexistence of strange attractors and quantum mechanics long time scales to be maintained in the non-vanishing of quantum commutators. This is feasible through a continuous dissipation strength controlling parameter  {$\epsilon$}.

\section{ Controlling Dissipation in the Lorenz System}
Before we study in detail the scaling properties of the extended Lorenz system ( we shall refer
to it as the $\epsilon$-Lorenz system)

\beqa
\dot{x} & = & \sigma (y - \epsilon x ) \nonumber \\
\dot{y} & = & x ( r - z) - \epsilon y \\
\dot{z} & = & x y - \epsilon b z \nonumber
\eeqa

lets analyze the stability properties of its critical points. On this issue we keep 
in mind that a similar model has been considered in the literature( \cite{howard,rob, gucken}) but with precisely fixed  control parameter
$ \epsilon = \frac{1}{\sqrt{\sigma r}} $,  in order that the limit $ r \rightarrow \infty $ becomes explicit. 
For the stability analysis of the critical points we follow \cite{sparrow} and \cite{gucken}.

The critical points of the system of eq. (2.1) are: $P_{0}: (x=y=z=0)$ the origin of the phase space, and $ P_{\pm}: (x_{\pm} = \pm \sqrt{b (r - \epsilon^{2})} , \ \ y_{\pm}= \pm\epsilon\sqrt{b(r -
\epsilon^{2})}, \ \ z_{\pm} = r-\epsilon^{2})$. In order that $P_{\pm}$ are real points, we must
have $ r > \epsilon^{2} $. Linearization of the system around $P_{0}$ gives

\beqa
\frac{d}{dt} \left( \begin{array}{c} \delta x \\ \delta y \\ \delta z \end{array} \right) \ \ = \ \ \left( \begin{array}{ccc} -\epsilon \sigma & \sigma & 0 \\ r & -\epsilon & 0 \\ 0 & 0 & -b\epsilon \end{array} \right) \left( \begin{array}{ccc} \delta x \\ \delta y \\ \delta z \end{array}\right) 
\eeqa

and the eigenvalue problem with

\vspace{4.0mm}

\beq
\left( \begin{array}{c} \delta x \\ \delta y \\ \delta z \end{array} \right) \ \ = \ \ 
e^{\lambda t } \left( \begin{array}{c} \delta x(0) \\ \delta y(0) \\ \delta z(0) \end{array} \right) 
\eeq
provides the characteristic Polynomial

\beq
P_{0}(\lambda) \ \ = \ \ - (\lambda + b \epsilon ) \ \left[ ( \epsilon \sigma + \lambda ) ( \epsilon + \lambda)
- \sigma r \right]
\eeq

It posesses two negative  and one positive eigenvalues. They correspond in turn, to a stable 2-dim. manifold  $ W^{s}_{0}$ with two attracting directions in $R^{3}$ as well as to an unstable 1-dim. manifold $ W^{u}_{0} $ with a single repulsive direction,  $ \forall \sigma>0,\ r > 
\epsilon^{2} $.

\beqa
\lambda_{1}&=&  -b\epsilon \ , \ \ \ \ \lambda_{2} \ = \ -\frac{1}{2} \left[ \epsilon(1+\sigma) + 
\sqrt{ \epsilon^{2} (1+\sigma)^{2} \ + \ 4 \sigma (r - \epsilon^{2}) } \right] \nonumber \\ 
\lambda_{3} &=& \frac{1}{2} \left[ \sqrt{\epsilon^{2}(1+\sigma)^{2} \ + \ 4 \sigma(r - 
\epsilon^{2}) } \ - \ \epsilon(1+\sigma) \right]
\eeqa 

Linearization, around the $P_{\pm}$ critical points, gives identical 
characteristic polynomials

\beq
-P_{\pm}(\lambda) \ = \  \lambda^{3} \ + \ \epsilon ( 1+ \sigma +b ) \lambda^{2} \ + \ b (r + \sigma \epsilon^{2}) \lambda \ + \ 2 \sigma b \epsilon \ (r - \epsilon^{2})
\eeq

Since all the coefficients are positive this cubic polynomial has at least one negative eigenvalue(attractive direction in $R^{3}$).

The bifurcation from 3-dim stable manifold (dim $W^{s}_{\pm}=3$) to 1-dim  stable 
 manifold   ( dim$W^{s}_{\pm}=1 $ ) occurs, when the real part of the pair of complex 
 conjugate roots
$ \lambda = \lambda_{R} \pm i \lambda_{I} $ crosses the value $ \lambda_{R}=0 $.  Simple algebra shows that $\lambda_{R}=0 $ is equivalent with the statement that
the constant term in $P_{\pm} $ in rel.(2.6) is the product of the coefficients of the linear and the quadratic terms.  Thus the critical value of r is given by,

\beq
r_{c}(\epsilon) \ = \ \epsilon^{2} \sigma \ \ \frac{\sigma + b + 3}{ \sigma - b - 1} 
\eeq

For $ r > r_{c}(\epsilon) $ the subcritical Hoph bifurcation leads to the aperiodic attractive orbits of
the famous Lorenz Strange attractor\cite{lorenz}.
From ref.(2.7) we get the interesting result that for small $\epsilon$ close to the non-dissipative integrable limit of $\epsilon=0$ the strange attractors appear for small values of  $ r > r_{c}(\epsilon) $.

In concluding we present   for the critical case of $\lambda_{R}=0$, \  
the other  roots of the  cubic polynomial $P_{\pm}$. 
The real root is given by

\beq
\lambda_{0}= -\epsilon(1 + \sigma + b) 
\eeq
whereas the imaginary component of the complex conjugate one is:
\beq
\lambda_{I} =  \frac{2  \epsilon^{2} \sigma(\sigma +1)}
{\sigma - b - 1}
\eeq 

$ \lambda_{I} $ represents the angular frequency of the orbit rotation 
around the critical points $P_{\pm}$.

In what follows we are going to investigate the scaling properties of the extended $\epsilon$-Lorenz system which will make transparent the $\epsilon$-dependence of $r_{c}(\epsilon) $ in rel. (2.7).

\section { Scaling Relations in the  Lorenz Model }

Firstly we rescale independently all the variables of the system  
$ \alpha, \beta, \gamma, \lambda \in R $ 

\beq
x= \alpha x^{\prime} \ , \ \  y = \beta y^{\prime}\ , \ \  z= \gamma z^{\prime} \ , \ \ 
 t = \lambda t^{\prime}
\eeq
We demand that in the transformed primed system a similar structure of constants appear in the equations of motion, i.e. of the  $ r , \epsilon, b, \sigma $ type  
( the dot represents time
derivative with respect to $t^{\prime}$),

\beqa
\dot{x^{\prime}} & = & \sigma^{\prime} (y^{\prime} - \epsilon^{\prime} x^{\prime} ) \nonumber \\
\dot{y^{\prime}} & = & x^{\prime} ( r^{\prime} - z^{\prime}) - \epsilon^{\prime} y^{\prime} \\
\dot{z^{\prime}} & = & x^{\prime} y^{\prime} - \epsilon^{\prime} b^{\prime} z^{\prime} \nonumber
\eeqa
 we obtain

\beq
\epsilon^{\prime} = \lambda \epsilon,\ \ \ b^{\prime}=b, \ \ \sigma^{\prime}=\sigma, \ \ \ r^{\prime}= \lambda^{2} r, \ \ 
\alpha= \frac{1}{\lambda}, \ \ \beta= \gamma = \frac{1}{\lambda^{2}}
\eeq

 The dynamical variables 
 $x,y,z$ are related to their primed ones as:

\beq
x(t)= \frac{1}{\lambda} \ x^{\prime}(\lambda t) \ , \ \  y(t)= \frac{1}{\lambda^{2}} y^{\prime}(\lambda t) \ , \ \ z(t) = \frac{1}{\lambda^{2}} z^{\prime} (\lambda t)
\eeq

We will show below that the established covariance property of the $\epsilon$-Lorenz system under the scalings (3.3 , 3.4 ) implies interesting
constraints for the dependence of the phase space coordinates $x,y,z$ on the time, the 
parameters r, $\epsilon, \sigma, $b as well as on the initial conditions $x_{0}, y_{0}, z_{0} $. 

Indeed the time evolution of the phase space coordinates is given by the exponential of
the Liouville operator ${\cal{L}}_{0}$, $( x_{0}=x(0), y_{0}=y(0), z_{0}=z(0))$,  
\beq
x(t;r,\epsilon,\sigma,b,x_{0},y_{0},z_{0}) \  = \  e^{t {\cal{L}}_{0}} \cdot x_{0}
\eeq

where

\beq
 {\cal{L}}_{0} = \sigma \left( y_{o} - \epsilon x_{0} \right) \partial_{x_{0}} + \left[ x_{0}(r - z_{0}) - \epsilon y_{0} \right] \partial_{y_{0}} + \left( x_{0} y_{0} - \epsilon b z_{0} \right) 
\partial_{z_0}
\eeq

Under the scaling (3.3-3.4) which we rewrite explicitly as 

\beq
t \rightarrow \frac{1}{\lambda}t \ \  , \ \   \sigma \rightarrow \sigma \ \ , \ \  b \rightarrow b \ \ , \ \  r \rightarrow \lambda^{2} r \ \  , \ \ \epsilon \rightarrow \lambda \ \epsilon
\eeq

as well as,

\beq
x_{0} \rightarrow \lambda x_{0} \ \ , \ \ y_{0} \rightarrow \lambda^{2} y_{0} \ \ , \ \ 
z_{0} \rightarrow \lambda^{2} z_{0}
\eeq

( $ \forall \lambda \in R $ ), 
it is straightforward to check that  $ {\cal{L}}_{0} $ scales as 

\beq
{\cal{L}}_{0} \ \rightarrow \ \lambda \  {\cal{L}}_{0}
\eeq
At last from rel.(3.5) we obtain that $ \forall t \in R $,

\beq
x( \frac{1}{\lambda}t  ; \ \lambda^{2} r , \  \lambda \epsilon , \ \sigma , \ b , \ \lambda x_{o} , 
 \ \lambda^{2} y_{o} , \  \lambda^{2} z_{o} )  \ =  \ \lambda \ x(t ; \ r  , \ \epsilon , 
 \ \sigma , \ b ,\  x_{o} ,\  y_{o} , \ z_{o} )
\eeq

and for the $y(t)$ and $z(t)$ coordinates we obtain respectively

\beq
y \ \rightarrow \ \lambda^{2} y \ \ , \ \ z \rightarrow \lambda^{2} z
\eeq

We note that the nondissipative Lorenz system ($ \epsilon = 0 $) satisfies scaling relation (3.10). When ($ \epsilon \neq 0$ )we impose  $ \lambda \cdot \epsilon =1 $ on the LHS of eq.(3.10) which implies that the whole r - $\epsilon$ plane is foliated by the continuous set of parabolas
\beq
\frac{r}{\epsilon^{2}} \ = \ r_{\cal{L}} \ = \ \mbox{constant}
\eeq

Each parabola corresponds to a different phase of the Lorenz system which is specified 
by the value of the Reynold's parameter r for which the parbola cuts the $\epsilon = 1 $
line . On the other hand if we fix a given parabola all of its points are related through
the scaling relations (3.3-3.4). Indeed the condition  $ \lambda \cdot \epsilon = 1 $ implies that:
\beq
x \left(t; \ r, \ \epsilon, \ \sigma, \ b, \ x_{0}, \ y_{0}, z_{0} \right)
\ = \ \epsilon \ x \left( \epsilon \ t ; \ \frac{r}{\epsilon^{2}}, \ 1 , \ \sigma , \ b , \ \frac{1}{\epsilon} x_{0}, \ \frac{1}{\epsilon^{2}} y_{0} , \ \frac{1}{\epsilon^{2}} z_{0} \right)
\eeq

Similarly for $y,z$ we have 

\beq
y \left(t; \ r, \ \epsilon, \ \sigma, \ b, \ x_{0}, \ y_{0}, z_{0} \right)
\ = \ \epsilon^2 \ y \left( \epsilon \ t ; \ \frac{r}{\epsilon^{2}}, \ 1 , \ \sigma , \ b , \ \frac{1}{\epsilon} x_{0}, \ \frac{1}{\epsilon^{2}} y_{0} , \ \frac{1}{\epsilon^{2}} z_{0} \right)
\eeq

\beq
z \left(t; \ r, \ \epsilon, \ \sigma, \ b, \ x_{0}, \ y_{0}, z_{0} \right)
\ = \ \epsilon^2 \ z \left( \epsilon \ t ; \ \frac{r}{\epsilon^{2}}, \ 1 , \ \sigma , \ b , \ \frac{1}{\epsilon} x_{0}, \ \frac{1}{\epsilon^{2}} y_{0} , \ \frac{1}{\epsilon^{2}} z_{0} \right)
\eeq

Equations (3.13-3.15) constitute the main result of the present work. It relates the time evolution of the extended $\epsilon$-Lorenz system $\epsilon \neq1 $ with the standard Lorenz one $ \epsilon=1 $. Indeed, by denoting with 
 $ r_{\cal {L}}$ the ratio $ r_{\cal{L}} = \frac{r}{\epsilon^{2}} $ we can read off eqs.(3.13-3.15) in two distinct but complementary ways.

Firstly, if we fix  $ r_{\cal L}$ we see that the r parameter of the 
$ \epsilon$-Lorenz system ( LHS  of eqs.(3.13-3.15)) ranges between the values  of 
 $ 0 < r <  r_{\cal L}$ when $ 0 < \epsilon < 1 $. It follows that the physics of all the points
 on the parabolas $ r =  r_{\cal L} \cdot \epsilon^{2} $ of the 
 ( r -  $\epsilon$ ) plane is the same.  
 
On the other hand if we fix r for the $\epsilon$-Lorenz system ( LHS of (3.13-3.15) ),
by varying $\epsilon$ $( 0 < \epsilon < 1 )$,  we are able to scan the region of the standard Lorenz system, for  $ r_{\cal L}, \ r < r_{\cal L} < \infty $ (see fig.1). 
\begin{figure}[thp]
\centering
\includegraphics[scale=0.6]{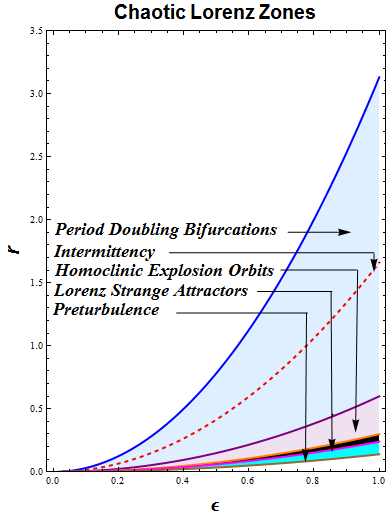}
\caption[]{Lorenz System Structure of Dynamics in r-$\epsilon$ plane }
\end{figure}



Some remarks are in oder: 
  
  By fixing r in the $\epsilon $-Lorenz system and taking the limit $ \epsilon \rightarrow 0 $ we recover the non-dissipative Lorenz system of eq.(1.4). 
It is integrable and describes the motion of a particle in an one dimensional anharmonic potential \cite{af1}( or the pendulum). We can still rescale this sytem by choosing $ \lambda = \frac{1}{\sqrt{r}}$ and we discover the infinite  r limit of the full Lorenz system, which has been studied in detail in ref.\cite{rob, shim}. 
In this limit the Lorenz attractor is degenerate into an 8-figure stable limit cycle  ( see also  in \cite{fowler}).
In the above mentioned works, the
 $ \frac{1}{r} $ correction has been studied and simple bifurcations of the limit cycle have been observed. In \cite{rob} an exhaustive search has been made by lowering the values of  r. It has been discovered  that for relative small values of  the 8-figure still survives somewhat deformed nevertheless up to $ r =300 $. 
 
In table 1 we summarize the behavior of the Lorenz system for various interval values of the Reynold's number $ r_{{\cal L}}$ {which
comprise the famous Feigenbaum Route to Chaos. In the present case of the extended $\epsilon$-Lorenz system with a (r, $\epsilon$ ) space of parameters they correspond to parabolic zones of equivalent dynamical systems, specified by the scaling relation  
 $ r=r_{\cal L} \epsilon^2 $. 
 In table 1 we depict two such characteristic equivalent Lorenz systems for $\epsilon=1$ and $ \epsilon =.01 $.  The Route to Chaos for each of them is characterised by the following distinct behaviours for different values of
 r for fixed $ \epsilon $ :

 For values of the  $r=r_{\cal L}$ with $\epsilon=1$  in the range 
$ (  30 < r < 313)  $ is composed of an infinite cascade
of period doubling bifurcations $(2,2^2 ,2^3 , \cdots , 2^{n} )$ which appear until a critical value of r , $ r^{c} = 30.1$ is reached below which  the strange attractors appear for r in the range
  $ 24.06 < r < 30.1$ . 
 At approximately the value of $r=r_{\cal L}=166.07$ we have the Intermittency scenario to Chaos which was first described by Pomeau and Maneville \cite{MP}. 
 For $ 13 < r < 24.06 $ an interesting ``preturbulence'' regime has been observed
 in the works of J.Yorke et.al.\cite{yorke}. For a complete and detailed description of the behaviour of the Lorenz system for all values of r one should look at the work of Sparrows who does it exhaustively\cite{sparrow}. All of the above are depicted in figure 1.

\begin{center}
\begin{table}
\caption{(r-$\epsilon)$\ \ Lorenz \ \ Roadmap \ \ to \ \ Chaos}
\begin{tabular}{||c||c|c|}
\hline
 & $r=r_{{\cal L}}$\ , $\epsilon=1$ & r , $\epsilon=0.01 $ \\\hline
Period \ Doubling \ \ Bifurcations & \ \ 59.5 $<$  r $<$ 313 \  & \ 0.00596 $<$ r $<$ 0.0313  \\\hline
Intermittency & \ \  r \ $= $\ 166.07  \ &  \   r \ $= $\ 0.0166  \\\hline
Orbits from Homoclinic Explosions &   30.1 $<$ r $<$ 59.5 \ & \ 0.0030 $<$ r $<$  0.00596 \\\hline
Strange \ \ \ Attractors &  \ \ 24.06 $<$ r $<$ 30.1 \  & \  0.0024  $<$ r $<$ 0.00301 \\\hline 
Preturbulence & \ $ \ 13.9 < r < 24.06 \ $ & \  $ 0.00139 < r < 0.0024 $ \\\hline
\end{tabular}
\end{table}
\end{center}

\newpage

\section { On the Scale Invariant Lorenz System }

In the last part of this work we introduce a form of the $\epsilon$-Lorenz system which
is by construction invariant under the scalings in rel.(3.3-3.4). To this end we 
define new independent and dependent variables( $ r > \epsilon^{2} $ ) :

\beq
\tau =\sqrt{r- \epsilon^{2}} t \ , \ \ X = \frac{x}{\sqrt{r-\epsilon^{2}}} \ , \ \ Y=\frac{y}{r-\epsilon^{2}} \ ,  \ \ Z = \frac{z}{r-\epsilon^{2}}
\eeq

which satisfy the system of eqns.
\beq
\dot{X}= \sigma ( Y - \zeta X ) \ , \ \ \dot{Y} = X ( 1 + \zeta^{2} - Z ) - \zeta Y  \ , \ \ 
\dot{\zeta}= X Y - b \zeta Z
\eeq
with $ \zeta = \frac{\epsilon}{\sqrt{r - \epsilon^{2}}}$  

The derivative ``dot'' is with respect to the new time $\tau$ .
Every parabola $ \frac{r}{\epsilon^{2}} = c $ in the r - $\epsilon $ plane
is determined by a fixed value of 
\beq
\zeta = \frac{1}{\sqrt{\frac{r}{\epsilon^{2}} - 1 }} 
\eeq
 or 
 \beq
 r \ = \ \left( 1 + \frac{1}{\zeta^{2}} \right) \epsilon^{2}
 \eeq
 By construction it is easy to see that the variables $\tau $, X,Y,Z are scale invariant under rel.(3.3-3.4).
We may also observe that in the system (4.2) the parameter $\zeta $ appears
in both places where the Reynolds number as well as the dissipation control
parameter $\epsilon $appeared in eq.(2.1) independently.

Repeating the critical point analysis, as before, and their stability we find that there are three critical points. the point $P_{o}=(0,0,0) $ with two attracting and one repelling directions with corresponding eigenvalues:

\beqa
\lambda_{1}&=& -b\zeta\ , \ \ \ \lambda_{2} = - \frac{1}{2} [ \zeta(1+\sigma) \ + 
\ \sqrt{ \zeta^{2}(1+\sigma^{2} +4 \sigma } ]\ , \nonumber \\
\lambda_{3} &=& \frac{1}{2} \ [ \sqrt{ \zeta^{2}(1+\sigma^{2} +4 \sigma) }\ - \ \zeta (1+\sigma) ]
\eeqa

as well as two symmetric , under the reflection $(x,y,z) \rightarrow (-x,-y, z)$critical points  $  P_{\pm} = ( \pm\sqrt{b} \ , \pm\zeta\sqrt{b}  \ , \ 1 )$

with characteristic polynomial 
\beq
P_{\pm}(\lambda) = \lambda^{3} + \zeta( 1 + b + \sigma ) \lambda^{2} + b \left[ 
 1 + \zeta^{2} ( 1 + \sigma) \right] \lambda + 2 b \zeta \sigma
\eeq

Once more the bifurcation condition for the appearance of the strange attractor is 
$ \zeta < \zeta_{c} $ where 
\beq
\zeta_{c}^{2} = \frac{\sigma - b -1}{(\sigma+1)(\sigma + b + 1)}
\eeq

For the standard values of $\sigma=10$ and $b=\frac{8}{3}$ we find 
\beq
\zeta_{c}^{2} = \frac{19}{11 \cdot 41} = \frac{11}{451}
\eeq
We note finally that the scale invariant form of the evolution eqs. (4.2) can be cast in  a form, which
exhibits the role of dissipation on the non-dissipative sector of the system.
In figure 2 we plot $r_{{\cal L}} = \frac{r}{\epsilon^{2}}$ ,i.e. the relation of $r_{{\cal L}}$ against $\zeta$. Each value of $\zeta$ determines a unique parabola in the r-$\epsilon$ parameter plane of the equivalent physics of the Lorenz system.

\begin{figure}[thp]
\centering
\includegraphics[scale=0.6]{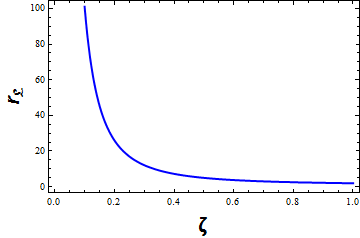}
\caption[]{ Lorenz $\zeta$ Dissipation Parameter}
\end{figure}

The Non-Dissipative part was defined in rel.(1.4) and corresponds to the $\zeta=0$ case : 
\beq
\dot{X}=\sigma Y , \ \ \  \dot{Y} = X(1 - Z) , \ \ \  \dot{Z} = X Y 
\eeq
We choose to work with the Cylinder intersecting with a Paraboloid as the two Nambu
``Hamiltonians'' among the whole set of SL(2,R) geometries  \cite{af1}.  
\beq
H_{1} = \frac{1}{2} \left[ Y^{2} + \left( 1 - Z \right)^{2} \right] , \ \ \ 
H_{2} = \sigma Z - \frac{X^{2}}{2}
\eeq
or 
\beq
\dot{\vec{X}} = \nabla H_{1} \ \times \ \nabla H_{2}
\eeq
For $\zeta=0$ the two surfaces $H_{1}, H_{2} $ are conserved 
\beq
H_{1}= H_{1}(t=0) \ \ , \ \ H_{2} = H_{2}(t=0) 
\eeq 
and they define by their intersection the trajectory of the system (4.9).  
When $\zeta \neq 0 $,\ $ H_{1}, H_{2} $ are not conserved. 
Still one can use in the place of  X, Y,  Z the variables   X and  $H_{2}$ in order to elucidate what
happens. For $\zeta \neq 0 $ we obtain \cite{shim}
\beq
\ddot{X} \ + \ ( 1 + \sigma ) \zeta \dot{X} \ + \ X \left[ \frac{X^{2}}{2} + H_{2} - 
\sigma \right] \ = \ 0 
\eeq
and
\beq
\dot{H_{2}} \ + \ \zeta \left[ b H_{2} \ - \ \sigma ( 1 - \frac{b}{2\sigma}) X^{2} 
\right] \ = \ 0
\eeq
We see now the crucial role of $ \zeta $ in the introduction of two terms: Firstly a friction term proportional to $ \dot{X}$ 
and secondly, a memory term in the anharmonic potential minimun. 
By eliminating $ H_{2} $ and using rel.(4.14) we obtain the Takeyama evolution memory term, which changes in a non-Marcovian  random manner the symmetry of the single well potential into the double well one \cite{takeya}.

 





\newpage
\section{ Conclusions-Open Problems}

The introduction of the $\epsilon $-Lorenz system, which controls the strength of the 
dissipative sector implies the existence of a weak (under-dissipated) and strong(over-dissipated) phases for the Lorenz system. They are separated in the 
$ r-\epsilon $ parameter space phase diagram by the critical line $ \epsilon=1$ which reproduces for different values of r the route to chaos of the 
Lorenz system, which is presented so lucidly
in the work of Sparrow\cite{sparrow}. We have demonstrated therein that a Lorenz system with weak dissipation
( small $\zeta$ values) is equivalent to the one  with large values of the Reynolds number r for the standard Lorenz system with $ \epsilon=1 $. 

The exploration of the scaling properties though, brought unexpected new information
about the $\epsilon$ Lorenz which smoothly joins to the $ \epsilon=0 $ integrable case. 
The scaling eqs.(3.3-3.4) show that we have very specific combinations of the time, the initial conditions and the parameters r and $\epsilon$ in each order in the Taylor expansion of the solution with respect to time, such that the scaling relation holds true. 
Also the bifurcation behaviour of the solution can be controled by the one parameter 
$ \zeta = \frac{\epsilon}{\sqrt{r - \epsilon^{2}}} $ in rel.(4.3) which parametrizes
the foliation of the r-$ \epsilon $ plane by one parameter family of parabolas. All
the points of each parabola are physically equivalent.  
Last but not least, the Nambu surfaces being the appropriate geometrical tool for the $\zeta=0 $ integrable case, are useful also for the small  $\zeta $ range which corresponds to the large r 
model ($\epsilon=1$) where we know that successive bifurcations of the figure 8 periodic limit cycle \cite{rob, shim} lead to the strange attractor configurations.
The Non-dissipative Limit is thus the gateway to Chaotic and turbulent Flows for the Lorenz system. 
Interestingly, it may be also the gateway to Quantum Chaos in a matrix formulation of the $\epsilon$-Lorenz system, where the expected presence of decoherence can become suppressed in an controllable way. This interesting possibility will be investigated in a separate work in the near future.

\section{Acknowledgments}
This research has been co-financed by the European Union (European Social Fund - ESF) and Greek national funds through the Operational Program "Education and Lifelong Learning" of the National Strategic Reference Framework (NSRF) - Research Funding Program: THALES. Investing in knowledge society through the European Social Fund. 
  E.G.Floratos acknowledges A.Bountis and S.Pnevmatikos for their kind hospitality, their interest and profitable discussions. 
  Last and foremost we are grateful to our dear friend and colleague, the late J.S.Nicolis, 
  whose passion for science in general, and dissipative chaotic systems in particular, has been an invaluable source of inspiration for both of us.

\end{document}